\documentclass[page-classic,linenumbers]{epl2} 
\title{Superconductivity by Berry connection
from many-body wave functions: a generalized Hartree-Fock approximation}
\shorttitle{} 

\author{Hiroyasu Koizumi}

\institute{Division of Quantum Condensed Matter Physics, Center for Computational Sciences, University of Tsukuba, Tsukuba, Ibaraki 305-8577, Japan         
}

\abstract{A fundamental revision of superconductivity theory that resolves the supercurrent carrier mass contradiction (the standard theory predicts it to be the effective mass but the London moment measurement indicates it to be the free electron mass) is presented, using a generalized Hatree-Fock approximation that takes into account a Berry connection from many-body wave functions.
 The new theory explains the pairing energy gap formation 
accompanying the superconductivity transition in the same manner as the standard theory, yet, provides  the free electron carrier mass in accordance with the London moment measurement.}
\begin{document}

\maketitle

\section{Introduction}

The standard theory of superconductivity is based on the BCS theory.
The origin of superconducting states is the electron-pair formation in this theory \cite{BCS1957}. 
By using the variational state vector
\begin{eqnarray}
|{\rm BCS}(\theta) \rangle =
\prod_{\bf k}(u_{\bf k}+ e^{i \theta}v_{\bf k}c^{\dagger}_{{\bf k} \uparrow} c^{\dagger}_{-{\bf k} \downarrow} )|{\rm vac} \rangle
\label{theta}
\end{eqnarray}
the BCS theory provides a way to calculate the energy gap formation due to the electron-pairing, where $|{\rm vac} \rangle$ is the vaccuum, $c^{\dagger}_{{\bf k} \sigma}$ is the creation operator for the conduction electron of effective mass $m^{\ast}$ with the wave vector ${\bf k}$ and spin $\sigma$.

The supercurrent is calculated as the linear response to the vector potential, yielding the carrier mass to have the effective mass $m^{\ast}$.
The carrier mass has been measured through the London moment for many materials \cite{Hirsch2013b}. The results always contradict the BCS theory, indicating that it is
the free electron mass $m_e$ instead of the effective mass $m^{\ast}$. 

 In the present theory, we will argue that the contradiction is resolved if $\theta$ in Eq.~(\ref{theta}) is defined differently by using
 the Berry phase from many-body wave functions \cite{koizumi2019,koizumi2021}.
 The phase $\theta$ arises in the standard theory the global $U(1)$ gauge symmetry breaking \cite{Anderson1958b,Nambu1960}, the number of particles in the superconductor is not definite; however, something corresponding to $\theta$ arises from spin-twisting itinerant motion of electrons, and inclusion of it resolves the carrier mass contradiction with keeping the particle number in the superconductor constant.

\section{Berry Connection for Many-Body Wave Functions}
\label{section2}

Let us first explain the Berry phase from many-body wave functions.
We consider the ground state wave function of a system of $N_e$ electrons,
$
\Psi ({\bf x}_1, \cdots, {\bf x}_{N_e},t)
$, 
where ${\bf x}_j=({\bf r}_j, \sigma_j)$ denotes the coordinate of the $j$th particle ${\bf r}_j$ and its spin $\sigma_j$.

A Berry connection associated with this wave function
can be defined as
 \begin{eqnarray}
{\bf A}^{\rm MB}_{\Psi}({\bf r},t)=-i \langle n_{\Psi}({\bf r},t) |\nabla_{\bf r}  |n_{\Psi}({\bf r},t) \rangle
\end{eqnarray}
where $|n_{\Psi}({\bf r},t) \rangle$ is defined as
\begin{eqnarray}
\langle  \sigma_1, {\bf x}_{2}, \cdots, {\bf x}_{N_e} |n_{\Phi}({\bf r},t) \rangle = { {\Psi({\bf r}, \sigma_1, {\bf x}_{2}, \cdots, {\bf x}_{N_e},t)} \over {|C_{\Psi}({\bf r} ,t)|^{{1 \over 2}}}}
\end{eqnarray}
where $|C_{\Psi}({\bf r} ,t)|$ is the normalization constant given by 
 \begin{eqnarray}
|C_{\Psi}({\bf r} ,t)|=\int d\sigma_1 d{\bf x}_{2} \cdots d{\bf x}_{N_e}\Psi({\bf r},\sigma_1, {\bf x}_{2}, \cdots)\Psi^{\ast}(\sigma_1, {\bf x}, {\bf x}_{2}, \cdots)
\end{eqnarray}

A more explicit expression for ${\bf A}^{\rm MB}_{\Psi}({\bf r},t)$ is given by
 \begin{eqnarray}
{\bf A}^{\rm MB}_{\Psi}({\bf r},t)&=&-i  \int d\sigma_1 d{\bf x}_{2} \cdots d{\bf x}_{N_e}
{ {\Psi^{\ast}({\bf r}, \sigma_1, {\bf x}_{2}, \cdots, {\bf x}_{N_e},t)} \over {|C_{\Psi}({\bf r} ,t)|^{{1 \over 2}}}}
 \nabla_{\bf r}
{ {\Psi({\bf r}, \sigma_1, {\bf x}_{2}, \cdots, {\bf x}_{N_e},t)} \over {|C_{\Psi}({\bf r} ,t)|^{{1 \over 2}}}}
\nonumber
\\
&=&{1 \over {\hbar\rho({\bf r}, t)}}{\rm Re} \left\{ 
 \int d\sigma_1  \cdots d{\bf x}_{N_e}
 \Psi^{\ast}({\bf r}, \sigma_1, \cdots, {\bf x}_{N_e},t)
  {\bf p}_{\bf r}
\Psi({\bf r}, \sigma_1, \cdots, {\bf x}_{N_e},t) \right\}
\nonumber
\\
\end{eqnarray}
Thus, ${\bf A}^{\rm MB}_{\Psi}({\bf r},t)$ can be identified as a velocity field generated by the wave function $\Psi$ multiplied by $m_e/\hbar$; it describes the effect on the electron with ${\bf r}_1$ by the interactions with other particles with ${\bf x}_2, \cdots, {\bf x}_{N_e}$ through the wave function they share. 
The same effect exists for other particles; thus, ${\bf A}^{\rm MB}_{\Psi}({\bf r},t)$ is the common velocity field to all the particles forming the wave field $\Psi$, which can be taken as  a Schr\"{o}dinger field.

Using $\Psi$ and ${\bf A}_{\Psi}^{\rm MB}$, we define another function $\Psi_0$ 
\begin{eqnarray}
\Psi_0 ({\bf x}_1, \cdots, {\bf x}_{N_e},t)=\Psi ({\bf x}_1, \cdots, {\bf x}_{N_e},t)\exp\left(- i \sum_{j=1}^{N_e} \int_{0}^{{\bf r}_j} {\bf A}_{\Psi}^{\rm MB}({\bf r}',t) \cdot d{\bf r}' \right)
\label{wavef0}
\end{eqnarray}
The Berry connection for this state is calculated to be zero
 \begin{eqnarray}
{\bf A}^{\rm MB}_{\Psi_0}({\bf r},t)=-i \langle n_{\Psi_0}({\bf r},t) |\nabla_{\bf r}  |n_{\Psi_0}({\bf r},t) \rangle =0
\end{eqnarray}
Thus, $\Psi$ can be decomposed into the velocity field zero part $\Psi_0$, and the nonzero velocity part $e^{i \sum_j\int_{0}^{{\bf r}_j} {\bf A}_{\Psi}^{\rm MB}({\bf r}',t) \cdot d{\bf r}'}$.

Usually, the case is considered where the common velocity field is zero in the ground state. However, we consider the case where it is not zero in the following.

\section{The Hartree-Fock approximation with non-trivial ${\bf A}_{\Psi}^{\rm MB}$}

Let us revisit the Hartree-Fock approximation, where an approximate wave function $\Psi$ is given by a Slater determinant of single-particle wave functions $\psi_{k_j}({\bf x}_\ell)$,
\begin{eqnarray}
 \Psi ({\bf x}_1,\cdots,  {\bf x}_{N_e})={ 1 \over \sqrt{N_e!}} 
 \left|
 \begin{array}{ccc}
 \psi_{k_1}({\bf x}_1) & \cdots &  \psi_{k_{N_e}}({\bf x}_1) \\
  \vdots &    \vdots &    \vdots \\
    \psi_{k_1}({\bf x}_{N_e}) &   \cdots &  \psi_{k_{N_e}}({\bf x}_{N_e})  \\
 \end{array}
 \right|
\end{eqnarray}

Usually, the basis functions are taken to be
\begin{eqnarray}
\phi_{k_i}({\bf x}_1)=\varphi_{\lambda_i}({\bf r}_1) \langle \sigma_1| \uparrow \rangle \mbox{ or } \varphi_{\lambda_i}({\bf r}_1) \langle \sigma_1|\downarrow \rangle
\end{eqnarray}
where $\varphi_{\lambda_i}({\bf r}_1)$ is a single-valued function of the coordinate ${\bf r}_1$, and $ \langle \sigma_1| \uparrow \rangle$ and $ \langle \sigma_1|\downarrow \rangle$ are up and down spin functions of the spin coordinate $\sigma_1$, respectively. 

Now, we consider the case where single-particle states are those of spin-twisting itinerant motion. In this case, the spin function is coordinate-dependent, and the requirement that $\phi_{k_i}({\bf x}_1)$ is a single-valued with respect to the coordinate ${\bf r}_1$ must be imposed.

The Hartree-Fock approximation has been developed by tacitly assuming ${\bf A}_{\Psi}^{\rm MB}=0$; in this case, the obtained approximate wave function is that for $\Psi_0$.  We express it as
\begin{eqnarray}
 \Psi_0 ({\bf x}_1,\cdots,  {\bf x}_{N_e})={ 1 \over \sqrt{N_e!}} 
 \left|
 \begin{array}{ccc}
 \tilde{\psi}_{k_1}({\bf x}_1) &   \cdots &   \tilde{\psi}_{k_{N_e}}({\bf x}_1) \\
  \vdots &    \vdots & \vdots \\
     \tilde{\psi}_{k_1}({\bf x}_{N_e}) & \cdots &   \tilde{\psi}_{k_{N_e}}({\bf x}_{N_e})  \\
 \end{array}
 \right|
\end{eqnarray}

If the electrons perform spin-twisting itinerant motion, $\Psi_0$ may be a multi-valued function of coordinates. In this cate, $\Psi_0$ is not a legitimate wave function since it is not single-valued with respect to the coordinates.

When $\Psi_0$ is multi-valued, we construct a single-valued $\Psi$ as
\begin{eqnarray}
 \Psi ({\bf x}_1,\cdots,  {\bf x}_{N_e})=\exp \left( i \sum_{j=1}^{N_e} \int_{0}^{{\bf r}_j} {\bf A}_{\Psi}^{\rm MB}({\bf r}',t) \cdot d{\bf r}' \right) \Psi_0
 \label{single-valued}
 \end{eqnarray}
This is the legitimate wave function; and this is the wave function we employ for the extended Hartree-Fock procedure presented in this work.
It is not a Slater determinant of single-particle wave functions; it contains many-body effects through the gauge field ${\bf A}_{\Psi}^{\rm MB}$.

The form of the wave function in Eq.~(\ref{single-valued}) and the fact that $\Psi_0$ is a currentless state read that the kinetic energy is a sum of the contribution from $e^{- i \sum_j\int_{0}^{{\bf r}_j} {\bf A}_{\Psi}^{\rm MB}({\bf r}',t) \cdot d{\bf r}'}$ and $\Psi_0$ \cite{Bohm1949}; the former is given by
 \begin{eqnarray}
\int { \hbar^2 \over {2 m_e}}{\bf A}_{\Psi}^2 \rho({\bf r}) d^3 r
\label{Supercurrent-E}
\end{eqnarray}
where $\rho({\bf r})$ is the electron number density.
Supercurrent is generated from this term. It is not the linear response current assumed in the standard theory \cite{BCS1957}. 

For convenience, let us define an angular variable $\chi$  by
\begin{eqnarray}
{ {\chi({\bf r})} \over 2}= \int^{{\bf r}}_0 {\bf A}_{\Psi}^{\rm MB}({\bf r}',t) \cdot d{\bf r}' 
\end{eqnarray}
It is a multi-valued function of ${\bf r}$ when the Berry connection is nontrivial.

We denote the total energy as $E[\chi]$, indicating that it is a functional of $\chi$. We can construct $\chi$ using $\tilde{\psi}_{k_i}$ obtained from the standard Hartree-Fock procedure by imposing the single-valued condition on
$
   \tilde{\psi}_{k_j}({\bf x}) \exp\left( i  { \chi \over 2} \right)
   \label{eqSingle}
$
and the conservation of the local charge \cite{Koizumi2021c}. We will demonstrate it using an example, later.

The appearance of $\chi$ generates a collective mode. 
Let us quantize it by treating $\Psi$ as a Schr\"{o}dinger field, and employing the canonical quantization procedure.
For this purpose, we consider the following Lagrangian
\begin{eqnarray}
{\cal L}=\langle \Psi |i \hbar \partial_t -H |\Psi \rangle=
i\hbar \langle \Psi_0 |\partial_t |\Psi_0 \rangle-\hbar \int r dr d \phi \rho { \dot{\chi} \over 2} -\langle \Psi |H |\Psi \rangle
\end{eqnarray}

The canonical conjugate momentum of $\chi$ is calculated as
\begin{eqnarray}
\pi_{\chi}={{\delta {\cal L}} \over {\delta \dot{\chi}}}=-{ \hbar \over 2} \rho
\end{eqnarray}

The canonical quantization condition is given by
$[\pi_{\chi}({\bf r},t), \chi ({\bf r}',t)]=-i \hbar \delta({\bf r}-{\bf r}')$, yielding
\begin{eqnarray}
[\rho({\bf r},t), { 1 \over 2} \chi ({\bf r}',t)]=i \delta({\bf r}-{\bf r}')
\end{eqnarray}

Next, we introduce the following boson field operators,
\begin{eqnarray}
\psi_{\chi}^{\dagger}({\bf r})=\sqrt{\rho({\bf r})}e^{-{ i \over 2}
\chi({\bf r})}, \quad \psi_{\chi}({\bf r})=e^{{ i \over 2}\chi({\bf r})}\sqrt{\rho({\bf r})}
\end{eqnarray}
They satisfy the following commutation relation
\begin{eqnarray}
[ \psi_{\chi}({\bf r}), \psi_{\chi}^{\dagger}({\bf r}')]=\delta({\bf r}-{\bf r}')
\end{eqnarray}

Then, we construct the following boson creation and annihilation operators
\begin{eqnarray}
B^{\dagger}_{\chi} =\int rdr d\phi \psi_{\chi}^{\dagger}({\bf r}),
\quad B_{\chi} =\int rdr d\phi \psi_{\chi}({\bf r})
\end{eqnarray}
that satisfy the following commutation relation
\begin{eqnarray}
[ B_{\chi}, B_{\chi}^{\dagger}]=1
\end{eqnarray}

Finally, we obtain the following number operator for the collective mode,
\begin{eqnarray}
\hat{N}_{\chi}=B_{\chi}^{\dagger} B_{\chi}, \quad \hat{N}_{\chi}|{N}_{\chi} \rangle
={N}_{\chi}|{N}_{\chi} \rangle
\end{eqnarray}
The number ${N}_{\chi}$ is the number of electrons participating in the collective mode described by $\chi$, and $|{N}_{\chi} \rangle$ denotes its eigenstate.

We can also construct a phase operator $\hat{X}$ that is conjugate with $\hat{N}_{\chi}$; it is defined through $
B_{\chi}^{\dagger} =\sqrt{\hat{N}_{\chi}}e^{-i\hat{X}}$ and  $B_{\chi} =e^{i\hat{X}}
\sqrt{\hat{N}_{\chi}}$.

The phase and number operators have the following commutation relation
\begin{eqnarray}
[e^{i\hat{X}}, \hat{N}_{\chi}] =e^{i\hat{X}}
\end{eqnarray}
thus, 
$e^{\pm i\hat{X}}$ are number changing operators that satisfy
$
e^{\pm i\hat{X}}|{N}_{\chi} \rangle =|{N}_{\chi} \mp 2 \rangle
$.
These number changing operators are crucial for the generation of the pairing gap with keeping the total number of  number electrons constant.

Note that since $\chi$ and $\rho$ are now operators,  $\chi$ and $\rho$ appear in Eq.~(\ref{Supercurrent-E}) should be regarded as expectation values of the corresponding operators. 
 
\section{An example for a two-dimensional system of electrons with a pairing interaction}

To investigate the situation where ${\bf A}_{\Psi}^{\rm MB} \neq 0$,
let us consider a two dimensional system with the following single particle Hamiltonian
\begin{eqnarray}
h=-{{\hbar^2} \over {2 m_e}}(\partial_x^2 +\partial_y^2)+U(r)
\end{eqnarray}
 For simplicity we assume that the potential $U$ depends only on $r$.

The coordinate part of the wave function is given as the product of an angular function and a radial function given by
\begin{eqnarray}
\varphi_{nm}(r, \phi)={1 \over \sqrt{2 \pi}} e^{ i m \phi} R_{n |m|}(r), \quad h\varphi_{nm}(r, \phi)=E_{n m}\varphi_{nm}(r, \phi)
\end{eqnarray}
where $x=r \cos \phi, y=r \sin \phi$, $m$ is an integer, $n$ is a natural number that denotes the number of nodes of the radial wave function, $R_{n |m|}(r)$, which depends on $m$ through $|m|$.  The energy is given by $E_{n m}$.

Usually, the wave functions
\begin{eqnarray}
\psi_{n m \uparrow}= \varphi_{n m}(r, \phi)\langle \sigma|\uparrow \rangle, \quad \psi_{n m \downarrow}=\varphi_{n m}(r, \phi)\langle \sigma|\downarrow \rangle
\end{eqnarray}
are used by adopting the coordinate independent spin functions $|\uparrow \rangle$
and $|\downarrow \rangle$.
However, we consider the following spin functions
\begin{eqnarray}
|\Sigma_a (\phi) \rangle={ 1 \over \sqrt{2}}(e^{-{i \over 2}\phi} \sin \zeta |\uparrow \rangle 
+e^{{i \over 2}\phi} \cos \zeta  |\downarrow \rangle ),
\quad
|\Sigma_b (\phi) \rangle={ 1 \over \sqrt{2}}(-e^{-{i \over 2}\phi} \sin \zeta  |\uparrow \rangle 
+e^{{i \over 2}\phi}  \cos \zeta|\downarrow \rangle )
\end{eqnarray}
and use the following wave functions,
\begin{eqnarray}
\tilde{\psi}_{n m a}= \varphi_{n m}(r, \phi)\langle \sigma|\Sigma_a  (\phi) \rangle, \quad \tilde{\psi}_{n m b}=\varphi_{n m}(r, \phi)\langle \sigma|\Sigma_b  (\phi) \rangle
\end{eqnarray}
Expectation values of the components of spin for $|\Sigma_a (\phi) \rangle$ are given by
\begin{eqnarray}
\langle \Sigma_a (\phi)|s_x|\Sigma_a (\phi) \rangle={ \hbar \over 2}\cos \phi \sin \zeta,
\langle \Sigma_a (\phi)|s_y|\Sigma_a (\phi) \rangle={ \hbar \over 2}\sin \phi \sin \zeta,
\langle \Sigma_a (\phi)|s_z|\Sigma_a (\phi) \rangle={ \hbar \over 2} \cos \zeta
\end{eqnarray}
It describing spin-twisting around the $z$-axis. For simplicity, we consider the case where $\zeta$ is constant in the following.
Expectation values of the components of spin for $|\Sigma_b (\phi) \rangle$ are given by
$\langle \Sigma_b (\phi)|{\bf s}|\Sigma_b (\phi) \rangle=-\langle \Sigma_a (\phi)|{\bf s}|\Sigma_a(\phi) \rangle$, thus,  it describes the spin-twisting motion with the spin direction opposite to $|\Sigma_a (\phi) \rangle$.

 To have the single-valued total wave function, we require that
\begin{eqnarray}
{\psi}_{n m a}=\tilde{\psi}_{n m a}e^{i { \chi \over 2}}, \quad
{\psi}_{n m b}=\tilde{\psi}_{n m b}e^{i { \chi \over 2}} 
\end{eqnarray}
are  single-valued functions of the coordinate.
A satisfactory $\chi$ should only depends on $\phi$ since the multi-valuedness of $\tilde{\psi}_{n m a}$ and $\tilde{\psi}_{n m b}$ arise from their $\phi$ dependence.

The total energy $E[\chi]$ depends on $\chi$ through ${{d \chi} \over {d \phi}}$; 
thus, the condition for an optimal $\chi$ that minimize the total energy is given by
\begin{eqnarray}
0={ {\delta E[\chi]} \over {\delta \chi}}=-\nabla \cdot { {\delta E[\chi]} \over {\delta \nabla \chi}}=-{\partial \over { \partial \phi}}{ {\delta E[\chi]} \over {\delta {{d \chi}\over {d \phi}}}}
\end{eqnarray}
This it the equation for the conservation of the local charge, and yields, $0={{d^2 \chi}\over {d \phi^2}}$. Its solution is 
$
\chi=A \phi +B 
$,
where $A$ and $B$ are constants. The constant $B$ merely generates a constant phase factor on the wave function.
The constant $A$ must be so chosen that $e^{{i \over 2}(A\pm1)\phi}$ is a single-valued function of the coordinate. This condition yields that $A$ is an odd integer.

Using the obtained $\chi$, the total energy is given by
\begin{eqnarray}
E[\chi]=\int dr d \phi  {{\rho(r)\hbar^2}\over {2 m_e}}
\left({ 1 \over 2}{{d \chi}  \over {d \phi}} \right)^2
+\sum_{\tilde{E}_{n m} \leq 0}2 \tilde{E}_{n m}
\end{eqnarray}
where $\tilde{E}_{n m}={E}_{n m}-E_F$ and $E_F$ is the Fermi energy; the factor $2$ appears due to the fact that both $\psi_{n m a}$ and $\psi_{n m b}$ are occupied.
This current carrying state is energetically higher than the currentless state  \cite{Bohm1949}. However, $e^{\pm i\hat{X}}$ make it possible to generate a lower energy state if the electron-pairing interaction exists.

Now, introduce the pairing interaction given by
\begin{eqnarray}
 H_{\rm pair}&=&\sum_{n, m, n',m'} V_{n m; n' m'} c^{\dagger}_{n m \uparrow} c^{\dagger}_{n -m \downarrow}
c_{n' -m' \downarrow} c_{n' m' \uparrow} 
\nonumber
\\
&=&\sum_{n, m, n',m'} V_{n m; n' m'} c^{\dagger}_{n m \uparrow} c^{\dagger}_{n -m \downarrow}  e^{i\hat{X}} e^{-i\hat{X}}
c_{n' -m' \downarrow} c_{n' m' \uparrow} 
\end{eqnarray}
where $c^{\dagger}_{n m \sigma}$ and $c_{n m \sigma}$ are creation and annihilation operators for $\psi_{n m \sigma}, \ \sigma=\uparrow, \downarrow$, respectively,
and $1= e^{i\hat{X}} e^{-i\hat{X}}$ is inserted in the second line.

The single-particle part of the Hamiltonian is given by
\begin{eqnarray}
 H_0=\sum_{n, m} \tilde{E}_{n m}( c^{\dagger}_{n m \uparrow} c_{n m \uparrow} +c^{\dagger}_{n m \downarrow} c_{n m \downarrow})
\end{eqnarray}

We employ a BCS type variational ground state 
\begin{eqnarray}
|{\rm Gnd}\rangle =\prod_{n,m}(u_{nm}+v_{nm}c^{\dagger}_{n m \uparrow} c^{\dagger}_{n -m \downarrow}
 e^{i\hat{X}})|{\rm Cnd} \rangle
 \label{eqGnd}
\end{eqnarray}
where $|{\rm Cnd} \rangle$ corresponds to $\Psi_0$, i.e., it is given by
\begin{eqnarray}
|{\rm Cnd} \rangle =\prod_{\tilde{E}_{nm}\leq 0} a^{\dagger}_{nm}b^{\dagger}_{n -m}|{\rm vac} \rangle
\end{eqnarray}
where $a^{\dagger}_{nm}$ and $b^{\dagger}_{n m}$ denote creation operators for single-particle states
 $\tilde{\varphi}_{n m a}$ and $\tilde{\varphi}_{n m b}$, respectively;
$u_{nm}$ and $v_{nm}$ are variational parameters that satisfy
$u_{nm}^2+v_{nm}^2=1$.

The operator
$c^{\dagger}_{n m \uparrow} c^{\dagger}_{n -m \downarrow}
 e^{i\hat{X}}$ acting on $|{\rm Cnd} \rangle$ annihilates two electrons in the occupied states of  $\tilde{\varphi}_{n m a}$ and $\tilde{\varphi}_{n m b}$, and creates electrons in
  $\psi_{n m \uparrow}$ and  $\psi_{n -m \downarrow}$ states; thereby the number of electrons in the collective mode is reduced by two.
 
The pairing energy gap is defined as
\begin{eqnarray}
 \Delta_{n m}&=&-\sum_{n, m, n',m'} V_{n m; n' m'} \langle {\rm Gnd}| e^{-i\hat{X}}c_{n' -m' \downarrow} c_{n' m' \uparrow} |{\rm Gnd}\rangle
 \nonumber
 \\
 &=&-\sum_{n, m, n',m'} V_{n m; n' m'} u_{n' m'}v_{n' m'}
\end{eqnarray}
Note that the it is calculated with keeping the particle number constant.

As in the BCS theory, we assume $V_{n m; n' m'}=-V$ if $|\tilde{E}_{n m}|, |\tilde{E}_{n' m'}| \leq \hbar \omega$ and zero otherwise with a cut-off energy $\hbar\omega_c$.
By following the BCS approximation \cite{BCS1957}, we obtain
\begin{eqnarray}
u_{n m}^2={ 1 \over 2} \left(1 +{{\tilde{E}_{n m}} \over {\sqrt{ \tilde{E}_{n m}^2+ \Delta^2}}} \right), \quad
v_{n m}^2={ 1 \over 2} \left(1 -{{\tilde{E}_{n m}} \over {\sqrt{ \tilde{E}_{n m}^2+ \Delta^2}}} \right), \quad
\Delta\approx 2 \hbar \omega_c e^{-{ 1 \over {N(0)V}}}
\end{eqnarray}
where $N(0)$ is the density of states at the Fermi energy.

The total energy is given by
\begin{eqnarray}
E_{\rm tot}&=&\int r dr d \phi  {{\rho(r)\hbar^2}\over {2 m_e}}
\left({ 1 \over 2}{{d \chi}  \over {d \phi}} \right)^2+2\sum_{m n} \tilde{E}_{n m} v_{n m}^2-{{\Delta^2} \over V}
\nonumber
\\
&=&\int r dr d \phi  {{\rho(r)\hbar^2}\over {2 m_e}}
\left({ 1 \over 2}{{d \chi}  \over {d \phi}} \right)^2-{1 \over 2}N(0)V\Delta^2
\end{eqnarray}
where the number of electrons in the collective mode is calculated as
\begin{eqnarray}
&&\int  rdr d \phi  \rho(r)=\sum_{ \tilde{E}_{n m}\leq 0}u^2_{n m}=
2N(0)\int_{-\hbar \omega_c}^0 
{ 1 \over 2} \left(1 +{{x} \over {\sqrt{ x^2+ \Delta^2}}} \right)dx
\nonumber
\\
&&=N(0)\left(\hbar \omega_c +\sqrt{\Delta^2}-\sqrt{ \hbar^2 \omega_c^2+ \Delta^2} \right)
\approx N(0)\left(\Delta-{ 1 \over {2 \hbar \omega_c}}\Delta^2 +{1 \over {8\hbar^3 \omega_c^3}}\Delta^4 \right)
\label{nonzero}
\end{eqnarray}
If the energy gap formation makes the current carrying state lower in energy than the currentless state, the superconducting state is realized.

A salient feature of the present theory is that
the supercurrent is generated by the collective mode whose kinetic energy is 
separately given with mass $m_e$. 
The number of electrons in this mode is nonzero when the pairing gap is formed as seen in Eq.~(\ref{nonzero}), thus the supercurrent carrying state and the nonzero pairing gap state coincide.

If we consider a more general setting by including the potential energy from the underlying ion lattice and effective field from other electrons, the spin-twisting itinerant motion occurs as the circular motion around a section of the Fermi surface of the metal \cite{koizumi2020}.

When a magnetic field exists, the vector potential from magnetic field ${\bf A}^{\rm em}$ appears in addition to the gauge field ${\bf A}^{\rm MB}_{\Psi}$. Then, the kinetic energy of the collective mode is given by
\begin{eqnarray}
E_{\chi}=\int d^3 r {{\hbar^2 \rho({\bf r})} \over {2 m_e}}
\left({ 1 \over 2}\nabla \chi+{e \over \hbar}{\bf A}^{\rm em}
\right)^2
\end{eqnarray}

Then, the supercurrent density is given by
\begin{eqnarray}
{\bf j}=-{{\partial E_{\chi}} \over {\partial {\bf A}^{\rm em}}}
=-{{e^2 \rho({\bf r})} \over {m_e}}
\left({ \hbar \over {2e}}\nabla \chi+{\bf A}^{\rm em}
\right)
\end{eqnarray}
This is a diamagnetic current that explains the Meissner effect. The presence of the angular variable $\chi$ yields the flux quantum ${ h \over {2e}}$. 

The velocity field associated with the above supercurrent is 
\begin{eqnarray}
{\bf v}_s
={{e} \over {m_e}}
\left({ \hbar \over {2e}}\nabla \chi+{\bf A}^{\rm em}
\right)
\end{eqnarray}

 The velocity field generated inside the superconductor by rotating it with an angular velocity ${\bm \omega}$ is given by
$
{\bf v}_{\rm rot}
={\bm \omega} \times {\bf r}
$.

Since supercurrent electrons move with the body to shield the electric field from the ion core, the condition ${\bf v}_s={\bf v}_{\rm rot}$ is satisfied. Then, the magnetic field 
$
{\bf B}^{\rm em}={{2m_e} \over e} {\bm \omega}
$
is generated inside of the superconductor. This formula has the free electron mass in accordance with the experimental results.
  
\section{Conclusion}
In the present theory, the supercurrent is not the linear response current, but the flow of electrons caused by the velocity field generated by the Berry connection from many-body wave functions. It predicts the carrier mass that agrees with the London moment measurement.




\end{document}